\documentclass[12pt,a4paper]{article}

\newif\ifpdf
 \ifx\pdfoutput\undefined
 \pdffalse
\else
 \pdfoutput=1
 \pdftrue
\fi

\ifpdf
\usepackage[pdftex,colorlinks]{hyperref}
\hypersetup{pdfstartview={Fit -32768},
  pdfpagemode=UseOutlines,
  plainpages=true,
  bookmarksnumbered=true
  }
\hypersetup{bookmarksopenlevel=2}

\fi

\usepackage{a4}
\usepackage{epsfig}
\usepackage[hmargin=2cm,vmargin=2cm,nohead]{geometry}

\newcommand{\degC}{$^{\circ}$C}

\begin{document}
\par
\begin{center}
\huge \textbf{Man-made climate change: \\
facts and fiction}
\end{center}

\begin{center}
\large  
\textbf{Michael Dittmar and Anne-Sylvie Nicollerat} \\[0.3cm] March 2004
\normalsize

\vspace*{1cm}

\textit{\textbf{Abstract}}
\end{center}

\vspace*{-0.3cm}

\noindent \textit{Important issues about climate change 
are summarized and discussed:}

\noindent \textit{A large body of evidence shows that
the world climate is getting warmer.
Climate models give a consistent explanation of this observation once
human-made emissions of greenhouse gases are taken into account.
Furthermore, the main source of greenhouse gases comes from
the burning of oil, gas and coal, mainly in the industrialized countries.
Without any change of behaviour, the possible
predicted consequences of this climate change
for the coming decades are very disturbing.
Today's (in)action's will have long-term consequences 
for the entire biosphere and the living conditions of many future 
generations.}

\noindent \textit{The combination of the various points related to the climate change leads to a final question:  
``For how long will Humanity continue to bury its head in the sand?''}

\vspace*{0.5cm}

\begin{itemize}
\item \textit{The average ``earth'' surface temperature is increasing} \cite{ipcc}.

Measurements of the earth's surface temperatures, studied for the entire planet and during the last 50--100 years
demonstrate that the earth is getting warmer.
The so-called average global temperature has increased since the late 19th century\footnote{This estimate is given 
with a 95\% confidence level.} 
by 0.6 $\pm$ 0.2 \degC. Observations since 1976
indicate a significant temperature increase of 0.17 \degC~every 10 years.
The observed temperature increase in the northern hemisphere is somewhat larger than that in the 
southern one. Obviously, when scientists talk about a world average, you will 
find regions where the observed temperature variations are below or above this ``world'' average. It is thus normal
to find regions with larger changes.  
For example, measurements performed for the last 30 years in Switzerland show an increase of between 1 and 2 \degC, corresponding
to an increase of about 0.5 \degC~every 10 years \cite{switzerland}.
It is not surprising either that some regions in the world even got colder.

In addition, many phenomena consistent with this warming have been reported from all around the planet \cite{warmingreports}:
\begin{itemize}
\item With the exception of Antarctica and a few other places, a rapid glacier melting is observed all around the world. 
Dramatic changes are seen for Alaska, Greenland, the Himalaya, the Arctic ice shelf and many mountainous regions.
The size of essentially all glaciers in the Alps has shrunk enormously, as easily observed 
even by hikers who visit the same glaciers in consecutive years \cite{glaciers}.
\item The winter season in many areas of the world became shorter, warmer, and the snow coverage decreased strongly. 
\item The mountain vegetation has changed according to the warmer temperature. 
\item More extreme weather conditions are being observed. However, the evidence does not seem to be 100\%
unbiased, as sensational weathers and their consequences are more extensively covered in the news today.   
\end{itemize}
 
Such changes are especially visible in regions with high mountains, where different temperature zones 
can be studied easily by everybody. In this case, it might be interesting to ask   
older people about their memories of winter snow coverage and temperatures. 

In contrast to these qualitative facts, quantifying these changes is quite complicated.
For example it is difficult to define a yearly world average temperature and  
it is even more difficult to measure this in a consistent way over something like the last 100 years.
For example one must be careful in using measurements in areas where the landscape has changed 
considerably over the last 100 years. Consequently not all data can be used for a ``world'' average and 
somebody has to decide whether a particular data set is usable.
Despite these difficulties, the community of scientists working on these questions, summarized regularly and in detail 
in the reports from the IPCC (Intergovernmental Panel on Climate Change),
came to the conclusion that the world average temperature increases much faster than at any given time period during the 
last few thousand years, as demonstrated with various techniques such as tree-rings and ice core studies.

\item \textit{The earth's atmosphere as a greenhouse?}

The sun warms up the earth through its radiation or more simply
through its light, allowing the presence of life on our
planet. An important part of this radiation is reflected back
resulting in an expectation of 255 degree Kelvin (-18 \degC) for the 
surface temperature.
However, an average temperature of 288 degree Kelvin (+15
\degC) is observed. The difference,  
well understood, originates from the various so-called greenhouse
gases in the atmosphere. The most important greenhouse
gases are water vapour (clouds), carbon dioxide (CO$_2$) and methane
(CH$_4$).
These gases are transparent to the visible light coming from the sun, 
but ``trap'', like in a greenhouse, some fraction of the  
infrared light which is reflected back from the earth
surface. Eventually some kind of 
equilibrium builds up, which defines an average world temperature.
It is this greenhouse effect that explains why the temperature on the planet earth 
is about 33 \degC~warmer than the one calculated without it \cite{greenhouse}.
%

Consequently, there must be a correlation between the concentration of greenhouse gases and the average temperature. 
Thus, an increase in the CO$_{2}$ concentration must result in a warmer average temperature.
Such correlations between the atmospheric carbone dioxide concentration
and the average temperature, covering a period of several thousand years,
have been reported from the detailed analysis of
ice cores from the Antarctica, Greenland and other places~\cite{icecores}.

\item \textit{Are the observed recent temperature changes larger than 
normal climate variations responsible for ice ages and warm ``interglacial periods"?}

It is a well known fact that the earth's climate has changed all the time 
and that glacial periods of a few thousand years 
are exchanged with warmer periods since millions of years. Such changes are known 
to be very slow, with periods of thousands of years, and 
appear to be strongly correlated with periodic changes of the average earth--sun distance  
with a known period of about 50 thousand years. Resulting temperature changes have been estimated to be  
up to 5 \degC~for about 50000 years. The corresponding fastest temperature changes, observed during the warming periods at the 
end of the ice ages, are found to be about 1.5--2 \degC~per 1000 years.
These ``fastest natural'' changes, corresponding to 0.15--0.2 \degC~for 100 years,  
are very different from the changes observed during the last 50 years (0.5 \degC) \cite{longtermchanges}.
Furthermore, the latest observations indicate an even faster climate change, showing 
that out of the 10 warmest years on record, 9 are found after 1990 and  
the three warmest years ever were 1998, 2002 and 2003. It is also interesting that the summer of 2003 in central Europe
has been so exceptionally warm that it convinced many people about the reality of the changing climate. 

A large variety of effects are now included in the complicated multiparameter climate models. Essentially all of these
models demonstrate that the observed temperature changes are correlated to the measured atmospheric 
CO$_{2}$ concentrations.
These models are adjusted to the data from previous periods, using for example 
the years from 1900 to 1980. Once this is done, the model predictions for the years 1980 to 2000
can be tested with the observed temperatures.
Within these models, different parameters may be varied, which allow for ``natural'' fluctuations 
and the ones expected from the CO$_{2}$ concentration in the atmosphere.
Using this approach one finds that the only consistent description of the observed 
temperature changes is obtained once the CO$_{2}$ concentration is taken into account.

These climate models are able to describe the physics of the 
overall temperature change and many regional phenomena such as precipitations. Nobody in the scientific community 
claims that these models can be used to describe all climate effects in detail.
For example, the overall increase of the earth's surface temperature and many regional variations in the northern hemisphere are 
nicely explained. In contrast, the temperature variations in the southern hemisphere (and especially in the Antarctic) and   
in the upper atmosphere above a height of 10 km are currently not too well reproduced \cite{modelfailures}.

\item \textit{What are the reasons for the observed increase of CO$_{2}$ in the earth's atmosphere?}

Large amounts of carbon atoms are bound in plants (biomass) and the so-called fossil fuels 
(oil, natural gas and coal). Basically, fossil fuels are accumulated very old biomass,
produced over hundreds of millions of years.
Today, these fossil fuels are burned rapidly, resulting mostly 
in CO$_{2}$ and a large amount of thermal energy. Some fraction of this thermal energy is transformed into
mechanical and electric energy, which is then used intensively in
industrialized countries.
These fossil fuel reserves are extensively used: during the last 50 years, about half of the known 
world oil reserves have been burned \cite{knownoil}.    

It it important to remember how much the industrialized countries and their associated ``way of living'' depend
on the use of fossil fuels, and especially oil. Thinking that about 50\% of the 
known world oil has been used during the last 50 years, we are immediately confronted with the question of how much 
more oil remains to be discovered. This is a difficult question, and many 
scientists specialized in geology believe that not much more oil will be discovered during the coming years.
Their most convincing argument is the fact that, 
despite very systematic searches using the newest techniques,
the largest amount of oil has been discovered 
roughly between 1960 and 1970 and that new findings have been declining since \cite{discoverypeak}. 

Furthermore, strong indications show that the rate of world oil production has more or less reached a maximum now. Consequently, 
the yearly world oil production will decrease soon, perhaps already within the next 5 to 10 years.
This prediction is based on the observation that the extraction from individual oil fields, large or small,
is following a Hubbert curve, some kind of bell shape curve \cite{hubbertcurve}. 
Once the maximum of the curve is passed, one observes a decrease in the
oil extraction for individual fields and for entire regions. 
It seems that even some unexpected discoveries of new oil fields similar to the ones in Saudi Arabia 
will postpone this unavoidable decline of oil extraction and consumption at most by a few years.  
Many warnings about this coming oil shock and its possible consequences
can be found in the literature \cite{endofoil}.

\item \textit{What will be the results of continuous economic growth and its associated increase of energy consumption?}
 
The correlation between todays CO$_{2}$ concentration in the atmosphere,
which increased from roughly 280 ppm (parts per million) about 100 years ago to about 370 ppm today, 
and the use of fossil fuels, mainly in the industrialized countries, is 
impressive. Following the current trends, often called ``business-as-usual scenario'',
a CO$_{2}$ concentration of 450 ppm is expected for 
the year 2030. This will increase further to 600 ppm or more at the end of the century \cite{futureco2}.

Some scientists disagree with this demand-based scenario.
Assuming that ``the industrial lifestyles'' all around the 
world will change dramatically once the near-by Hubbert oil production peak is reached, they claim that this  
``end-of `` the oil age will also imply that the remaining 
coal can simply not be burned fast enough to achieve CO$_{2}$ concentrations of 600 ppm \cite{oilco2limits}. 

In 1999, the world population of 6 billion people burned fossil fuels and biomass corresponding to a 
world production of roughly 23000 million tons of CO$_{2}$ \cite{co21999}. 
This corresponds to about 3.8 tons of CO$_{2}$ per year and per person.  
The industrialized countries (or the rich OECD countries), about a 20\% of the world population, produced 
almost 70\% of the total world-wide CO$_{2}$, corresponding to 
about 12.3 tons per person. Out of these, the United States contributed 
a total of 5500 million tons per year (roughly 20 tons per person). This can be compared with some other CO$_{2}$ production data 
per capita and per year
from other industrialized countries: 14.4 tons (Canada), 9.7 tons Germany and 5.7 tonnes in Switzerland.
This should be compared with some less industrialized large countries, which have much smaller per capita CO$_{2}$ production:   
2.3 tons in China, 1.1 tons in India, 0.7 tons in Pakistan and 1.8 tons in Brazil. 
A complete list of per capita CO$_{2}$ production for all countries can be found here~\cite{co21999}. 
These numbers clearly demonstrate who is to blame for the observed climate change. Furthermore,  
we must know that the CO$_{2}$ currently produced
will remain for at least 100--200 years in the atmosphere.
Thus, even assuming that today's fossil-fuel burning could from now on be kept at a constant level,
the CO$_{2}$ concentration in the atmosphere will increase during the next 20--30 years to at least 450 ppm.

Using some modest assumptions about future CO$_{2}$ emissions, 
the climate models can be used to predict the resulting average world temperature, which  
is predicted to rise by another 2 to 5 \degC~within the next 50--100 years \cite{ipccprediction}.
This can be compared with the 2003 record summer temperatures in central Europe, which 
were measured to be 2--3 \degC~higher than the long-term average European summer temperatures.

\item \textit{Are the models, which predict consequences from the
expected CO$_{2}$ increase, optimistic?}

Nobody knows what will happen exactly and the modeling errors are large. Thus 
even the often quoted upper limit on the possible temperature 
increase of 5 \degC~might be too optimistic, and
even catastrophic scenarios are possible. 
For example, a temperature increase by about 8 \degC~
could result in the melting of the Antarctic ice shield, with 
a rise of the sea water level by several meters, forcing billions of people into migration with (un?)imaginable
consequences. 
Such temperature changes could also result in the melting of the permafrost region in the northern hemisphere, 
were a large amount of methane, known as a very powerful greenhouse gas, might be released. 
Consequently, the temperature might increase even faster and the possibility 
that life, as we know it, could become impossible for a large part of the planet can not be excluded.

\item \textit{How should humanity react?}

The answer to this question is more speculative and only reflexion
paths can be given here.
However, given all the known facts, the various extreme 
possibilities and the uncertainties of today's climate models into account,
it seems that the time for thinking is over and that world wide actions against the man-made climate change 
have to be started now. 

Thinking in a rational way, humanity should apply standards, that are used for other domains in our society!
For example, the introduction of new medicines for public use requires some 
type of safety proof, showing that there are no harmful secondary
effects. Applying a similar standard
would mean in this case that humanity has first to show that 
the consequences for the world climate are harmless, before the burning
of the fossil fuels can be continued.

Perhaps humankind should simply not be allowed to touch the 
climate, as long as the consequences are not understood. 
We can imagine a similar situation to that of parents who,
instead of trying to explain the dangers of electricity to small children, simply  
shout: ``Do not touch''!

\item \textit{The Kyoto Protocol (1997) \cite{kyotoref}.}

It is well known that the Kyoto Protocol, which tries to limit 
the CO$_{2}$ production of the industrialized countries to their emission values from 1990,
would only result in a marginal change of the world CO$_{2}$ production, 
and that the CO$_{2}$ concentration in the atmosphere would still increase
to about two times the pre-industrial level.

Nevertheless, once accepted, the Kyoto Protocol would demonstrate for the first time in human history that  
a minimal world-wide agreement for a major policy change would have been achieved
and more drastic measures could be imagined for the near future.

Unfortunately the Kyoto Protocol can now be considered as basically dead, as 
the current Bush government of the United States (joined recently by the Russian government) has decided to ignore it.
Furthermore, the entire CO$_{2}$ problem seems to be an irrelevant topic in the current US election campaign
and no near future changes in the US policy can be envisaged. 
On the other hand, Europe is not doing better. Although it ratified
the Kyoto protocol, it is failing to apply it and raised its global
CO$_2$ emissions in the last years.
Moreover we note that some of the most urgent problems of the
industrial societies like unemployement seem to be related to 
the missing economic growth. Thinking only in traditional terms, getting back to economic growth implies 
an increase in fossil fuel consumption \cite{cars}.

\item \textit{What needs to be done? What can be done?} 

Well, this is the domain where fiction, fantasy, dreams and nightmares 
have their place. 

Before proposing answers to these questions, let us summarize the issues explained so far:
\begin{enumerate}
\item The world climate is getting warmer.
\item Climate models show that the burning of oil, gas and coal in the industrialized countries
is responsible for the climate change.  
\item The expectations for the near future are very disturbing and many catastrophes are 
highly probable.
\item Today's (in)action's will have long-term consequences for the entire biosphere and the living conditions of many future 
generations.
\end{enumerate}

Only a small minority of people think about or argue for a radical change. 
These people basically demand that the rich industrial countries and their population reduce their use of fossil energy 
in such a way that the overall CO$_{2}$ production decreases as quickly as possible. 

The following example can be considered as a moderate radical proposal
for the next 25 years (!).
To limit the damage we could ask the yearly world CO$_{2}$ emission to be reduced so that  
the world emissions would be a factor of 2 smaller in the year 2025.
This could be achieved if the yearly world CO$_{2}$ production were reduced by about 2\% each year.

Can this be done, knowing that the world population is still increasing by roughly 70--80 million (about 1.3\%) every year?

A possible way would be to require the consumption in the rich countries to 
be reduced such that their CO$_{2}$ production per capita will match the ones from countries like India 
within the next 25 years, while 
the so-called developing countries would need to stabilize their current per capita consumption.
This scenario would require that the United States, central Europe and other industrialized countries   
reduce the fossil fuel burning by roughly 
10\% per year, and for the next 25 years.

Such a policy would certainly result in the reduction of some material comfort, not only in the richer countries. 
Parts of this CO$_{2}$ reduction can perhaps be compensated by a better overall efficiency 
and by the development of renewable energies.
However, an overall drastic reduction in the consumption of material goods 
appears to be unavoidable in this scenario. Such a policy 
could be enforced by requiring that
the largest consumers within each society had to reduce more dramatically 
than the ones.
%

Many people argue in favour of wait and see, or hope 
that the observations are not significant and that the 
models and their predictions are wrong. 
These people are joined by another large group of people who think that nothing can be done anyway,   
as humans are stupid (except themselves of course).

Another group of people, which live mainly in the ``rich'' countries,
even  naively
hope that climate change will have positive effects on their region and their life.
This is a very dangerous and selfish attitude.

Finally a large majority of the world population, either in the rich or in the poor 
regions, is already fully occupied with everyday's 
more or less serious survival problems.

Unfortunately, ignoring that the climate climate change is taking place 
with its possible consequences will not make it disappear and, wanted or not,  
today's fossil fuel burning mainly by the rich countries
is a serious problem for the whole of humanity.

Given the various issues related to the climate change, can we
continue to bury our heads in the sand~? \\
For the ones who feel powerless it might be good to remember: ``\textit{Even a one
thousand mile path starts with one step}''~\cite{end}.

\end{itemize}


\begin{thebibliography}{99}
\bibitem{longtermchanges} {Long-term temperature changes are discussed in \cite{ipcc}, section 2.4.3.}
\bibitem{modelfailures}{It is certainly wrong to use a particular model
in a region where the authors do not claim its validity. But doing this to 
claim that the entire model must be wrong goes beyond scientific honesty.
Some discussion about the validity of climate models can be found in \cite{ipcc}, section 8.10.3.}
\bibitem{futureco2} {The expectations for the projected climate change are summarized 
in the IPCC report \cite{ipcc}, section 9.3.}
\end{thebibliography}
\end{document}